# Sensitivity of principal Hessian direction analysis

**Luke A. Prendergast**

*La Trobe University*
*Dept of Mathematics and Statistics,*
*La Trobe University, VIC 3086, Australia.*
*e-mail:* luke.prendergast@latrobe.edu.au

**and**

**Jodie A. Smith**

*La Trobe University*
*Dept of Mathematics and Statistics,*
*La Trobe University, VIC 3086, Australia.*
*e-mail:* ja12smith@students.latrobe.edu.au

**Abstract:** We provide sensitivity comparisons for two competing versions of the dimension reduction method principal Hessian directions (pHd). These comparisons consider the effects of small perturbations on the estimation of the dimension reduction subspace via the influence function. We show that the two versions of pHd can behave completely differently in the presence of certain observational types. Our results also provide evidence that outliers in the traditional sense may or may not be highly influential in practice. Since influential observations may lurk within otherwise typical data, we consider the influence function in the empirical setting for the efficient detection of influential observations in practice.

## Contents

## 1. Introduction

Dimension reduction methods have increased in popularity in recent times due to an abundance of high-dimensional data. The increased acceptance of such methods gives rise to the need for further understanding with regards to the sensitivity of the associated estimators. For some dimension reduction methods, a consequence of this is the lack of diagnostics that can be used to detect influential observations. The purpose of this paper is to compare the sensitivity of two related, yet competing, dimension reduction methods and provide an influence diagnostic that is useful in practice.

Consider a univariate response variable $Y$ and $p$-dimensional predictor vector $\boldsymbol{X}$. In the regression setting, when $p$ is large it may be difficult to visually determine the complex structure relating $Y$ and $\boldsymbol{X}$ due to our own inability to visualize data in more than a few dimensions. As such, dimension reduction methods that seek to reduce the dimension of $\boldsymbol{X}$ without loss of important regression information are highly valued.

Here we examine the multiple-index model

$$Y = f(\mathcal{B}^\top \boldsymbol{X}, \varepsilon) \tag{1}$$

with $\mathcal{B} = [\boldsymbol{\beta}_1, \ldots, \boldsymbol{\beta}_K]$ where $\boldsymbol{\beta}_k$ $(k = 1, \ldots, K)$ are unknown $p$-dimensional column vectors, $\varepsilon$ is the error term with $\varepsilon \perp\!\!\!\perp \boldsymbol{X}$ (where $\perp\!\!\!\perp$ will denote independence throughout), $E(\varepsilon) = 0$ and $f$ is the unknown link function. If we let $\boldsymbol{\Gamma} = [\boldsymbol{\gamma}_1, \ldots, \boldsymbol{\gamma}_K]$ denote an arbitrary basis for $\mathcal{S} = \text{span}(\boldsymbol{\beta}_1, \ldots, \boldsymbol{\beta}_K)$, then dimension reduction without loss of information can be achieved by replacing $\boldsymbol{X}$ with $\boldsymbol{\Gamma}^\top \boldsymbol{X}$ when $K < p$. Li [13] calls $\mathcal{S}$ the effective dimension reduction (e.d.r) space and we will follow the lead of Cook [5] in assuming that $\mathcal{S}$ is a central subspace in that it is defined at its minimum dimension.

Many dimension reduction methods have been recently proposed that seek to identify $\mathcal{S}$ without prior knowledge of $f$ and only mild distributional conditions for $\boldsymbol{X}$. These include Sliced Inverse Regression (SIR, [13]), Sliced Average Variance Estimates (SAVE,[6]), SIRII [14], Principal Hessian Directions (PHD,[15]) and Minimum Average Variance Estimation (MAVE,[22]) to name a few.

Gather *et al.* [10; 11] show that, at the sample level, SIR can fail in the presence of just one 'bad' observation; a finding supported by way of the influence function by Prendergast [18; 19]. Prendergast [20] provided similar results via the influence function for SAVE and SIRII and showed that either of these methods or SIR may be the preferred choice, from a sensitivity standpoint, with respect to certain types of observations. Lue [17] introduced a trimming algorithm for one version of PHD that iteratively trimmed observations and was shown to work well under simulations of some perturbed models.

Despite the fact that two different versions of PHD were introduced by Li [15], there has been little in the way of developing sensitivity comparisons between them. Cook [4] notes that one of these versions may be preferable when the underlying model incorporates strong linear trends. The first purpose of this paper is to analyze and compare the sensitivity of these methods at the model. This allows for a deeper understanding into the detrimental effect that certain observational types may have in practice and allows us to explore the differences in the methods when dealing with such observations. As a consequence of such analyses, the second purpose of this paper is to introduce influence measures that can detect influential observations in practice.

## 2. Principal Hessian directions

Of the many recently proposed dimension reduction procedures, principal Hessian directions (PHD) is perhaps the most intuitive extension of existing methodology. Though the method was developed by Li [15] using Stein's Lemma [21], PHD is strongly related to Ordinary Least Squares (OLS) regression. Let $\boldsymbol{X} \sim N_p(\boldsymbol{\mu}, \boldsymbol{\Sigma})$ and suppose that the model given in (1) holds with $K = 1$. It can be shown that (See [2], [3], and [16]), under these conditions, where $\mu_y = E(Y)$, $\boldsymbol{\Sigma}_{\boldsymbol{x}y} = E\{(Y - \mu_y)(\boldsymbol{X} - \boldsymbol{\mu})\}$ , and $\boldsymbol{\Sigma}^{-1}\boldsymbol{\Sigma}_{\boldsymbol{x}y}$ denotes the OLS slope vector,

$$\boldsymbol{\Sigma}^{-1}\boldsymbol{\Sigma}_{\boldsymbol{x}y} \in \mathcal{S}. \tag{2}$$

Hence, in the single-index case where $K = 1$ for the model given in (1), OLS may be employed to derive a basis for $\mathcal{S}$ when the predictor variable is normally distributed. An exception to this is when $\boldsymbol{\Sigma}^{-1}\boldsymbol{\Sigma}_{\boldsymbol{x}y}$ in (2) is $\mathbf{0}$ in which case, whilst the OLS direction is trivially an element of $\mathcal{S}$, the direction itself does not provide a basis for $\mathcal{S}$.

Let $\boldsymbol{X} \sim N_p(\boldsymbol{\mu}, \boldsymbol{\Sigma})$ and denote $\mu_y = E(Y)$ and $\boldsymbol{\Sigma}_{y\boldsymbol{x}\boldsymbol{x}} = E\{(Y - \mu_y)(\boldsymbol{X} - \boldsymbol{\mu})(\boldsymbol{X} - \boldsymbol{\mu})^\top\}$. With the application of Stein's Lemma [21], Li [15] showed that the average Hessian matrix of $E(Y|\boldsymbol{X})$ is given as

$$\overline{\mathrm{H}}_{\boldsymbol{x}} = \boldsymbol{\Sigma}^{-1}\boldsymbol{\Sigma}_{y\boldsymbol{x}\boldsymbol{x}}\boldsymbol{\Sigma}^{-1} \tag{3}$$

where the eigenvectors corresponding to nonzero eigenvalues of $\overline{\mathrm{H}}_{\boldsymbol{x}}$ are elements of $\mathcal{S}$. Li also noted that adding a linear function of $\mathcal{B}^\top\boldsymbol{X}$ to $Y$ does not change $\overline{\mathrm{H}}_{\boldsymbol{x}}$ so that an alternative definition is

$$\overline{\mathrm{H}}_{\boldsymbol{x}} = \boldsymbol{\Sigma}^{-1}\boldsymbol{\Sigma}_{r\boldsymbol{x}\boldsymbol{x}}\boldsymbol{\Sigma}^{-1} \tag{4}$$

where $\boldsymbol{\Sigma}_{r\boldsymbol{x}\boldsymbol{x}} = E\{r(Y,\boldsymbol{X})(\boldsymbol{X} - \boldsymbol{\mu})(\boldsymbol{X} - \boldsymbol{\mu})^\top\}$ and $r(Y,\boldsymbol{X})$ is the OLS residual function.

The original PHD methods estimated the matrix $\overline{\mathrm{H}}_{\boldsymbol{z}}$ based on $\boldsymbol{Z} = \boldsymbol{\Sigma}^{-1/2}(\boldsymbol{X} - \boldsymbol{\mu})$ which provides an orthonormal basis for $\boldsymbol{\Sigma}^{1/2}\mathcal{S}$. Re-transformation using $\boldsymbol{\Sigma}^{-1/2}$ could then be utilized to provide a basis for $\mathcal{S}$. However, the eigenvectors based on non-zero eigenvalues of $\overline{\mathrm{H}}_{\boldsymbol{x}}$ provide an orthonormal basis for $\mathcal{S}$ and, as such, all further reference throughout this paper to the PHD methods will be concerning estimation of $\overline{\mathrm{H}}_{\boldsymbol{x}}$.

## 3. Perturbation analysis in the dimension reduction setting

Consider an arbitrary distribution function $F$ and define the contamination distribution, with respect to $F$ and contaminant point $w$, to be $F_\epsilon = (1-\epsilon)F + \epsilon\Delta_w$ where $0 < \epsilon < 1$ and $\Delta_w$ is the Dirac measure putting all of its mass at $w$. Consider a statistical estimator with functional $t$ defined at $F$ and $F_\epsilon$. The influence function [12] for $t$ at $F$ is defined to be

$$\mathrm{IF}(t, F; w) = \lim_{\epsilon \downarrow 0} \left\{ \frac{t(G_\epsilon) - t(G)}{\epsilon} \right\} = \left. \frac{\partial\, t(G_\epsilon)}{\partial \epsilon} \right|_{\epsilon=0}. \quad (5)$$

The influence function approximates the relative influence of an observation $w$ from a large sample generated from $F$ on the estimator $t$.

Perturbation analysis in dimension reduction seeks to study the effect of small perturbations on detecting a correct basis for $\mathcal{S}$. Let $b_k$ $(k = 1, \ldots, K)$ denote the functional for an e.d.r. direction estimator with, for an arbitrary distribution $F$, $\|b_k(F)\| = 1$ and $b_i(F)^\top b_j(F) = 0$ $(i \neq j)$. Also, let $(Y, \boldsymbol{X}) \sim G$ such that the model in (1) is satisfied and $\mathrm{span}\{b_1(G), \ldots, b_K(G)\} = \mathcal{S}$ such that $b_1(G), \ldots, b_K(G)$ provide a basis for $\mathcal{S}$.

In the dimension reduction setting define the contamination distribution function as

$$G_\epsilon = (1-\epsilon)G + \epsilon\Delta_{(y_0, \boldsymbol{x}_0)} \quad (6)$$

where $0 < \epsilon < 1$ and $\Delta_{(y_0,\boldsymbol{x}_0)}$ is the Dirac measure putting all of its mass at the point $(y_0, \boldsymbol{x}_0) \in \mathbb{R}^{p+1}$. Let $\mathcal{S}_\epsilon = \mathrm{span}\{b_1(G_\epsilon), \ldots, b_K(G_\epsilon)\}$ be the equal-dimension perturbed equivalent of $\mathcal{S}$.

Since the basis for $\mathcal{S}$ is of primary relevance, a perturbation analysis seeking changes in $\mathcal{S}_\epsilon$ should not simply compare $\mathcal{S}$ and $\mathcal{S}_\epsilon$ column by column. Following the lead of Bénasséni [1], one approach is to study the angle between each $b_k(G_\epsilon)$ and its projection onto $\mathcal{S}$. In noting that many measures of angle are insensitive to small perturbations, Bénasséni introduced a measure between spans that utilized the average sine of the angle between each element of one basis and its projection onto the space spanned by the other. Bénasséni then also derived the influence function for this measure based on eigenvector subsets of the covariance matrix estimator.

Prendergast [19] utilized Bénasséni's measure for a sensitivity analysis of SIR using the influence function. Prendergast [20] extended this result to include the methods SAVE and SIRII and provided useful sensitivity comparisons between these methods and SIR. For a given $(y_0, \boldsymbol{x}_0)$, the influence function for this measure is simply the negative average of the sine of the angle between each perturbed direction and its projection onto the unperturbed space relative to $\epsilon \downarrow 0$. Hence, the sine of this angle can be seen as a relative increase in sine due to an $\epsilon$-perturbation. We now provide a formal definition of the Relative Increase in Sine with respect to the $k$th e.d.r. direction estimator.

**Definition 3.1.** *Using the notation defined above, let $\theta_{\epsilon,k}$ denote the angle between $b_k(G_\epsilon)$ and its projection onto $\mathcal{S}$. The Relative Increase in absolute*

*Sine (RIS) for the kth direction is defined to be*

$$\text{RIS}(b_k, G; y_0, \boldsymbol{x}_0) = \left| \lim_{\epsilon \downarrow 0} \frac{\sin(\theta_{\epsilon,k})}{\epsilon} \right|$$

*at G.*

*Remark* 3.1. Let $s$ denote the statistical functional such that, at an arbitrary distribution $F$, $s(F) = \sin(\theta_F)$ where $\theta_F$ is the angle between $b_k(F)$ and its projection onto $\mathcal{S}$. Then, with $\theta_{\epsilon,k}$ defined as in Definition 3.1, and since $\sin(\theta_{0,k}) = 0$, then

$$\text{RIS}(b_k, G; y_0, \boldsymbol{x}_0) = |\text{IF}(s, G; y_0, \boldsymbol{x}_0)| \, .$$

*Remark* 3.2. There is a strong link between the RIS and the influence functions for SIR, SAVE and SIRII considered by [19; 20] in that they are equal to

$$-\frac{1}{K} \sum_{k=1}^{K} \text{RIS}(b_k, G; y_0, \boldsymbol{x}_0)$$

under the appropriate conditions for which they were defined.

Assume $\theta_{\epsilon,k} \in [-\pi, \pi]$. The RIS has the following properties:

i) When $\theta_{\epsilon,k} = \pm\pi$ or $\theta_{\epsilon,k} = 0$ then $b_k(G_\epsilon) \in \mathcal{S}$ and $\text{RIS}(b_k, G; y_0, \boldsymbol{x}_0) = 0$.
ii) When $\theta_{\epsilon,k} = \pm\pi/2$ then $b_k(G_\epsilon) \perp \mathcal{S}$ and $\text{RIS}(b_k, G; y_0, \boldsymbol{x}_0) = \infty$.
iii) When $b_k(G_\epsilon)$ is rotated away from $\mathcal{S}$, $\text{RIS}(b_k, G; y_0, \boldsymbol{x}_0)$ increases.
iv) When $b_k(G_\epsilon)$ is rotated towards $\mathcal{S}$, $\text{RIS}(b_k, G; y_0, \boldsymbol{x}_0)$ decreases.

Closed-form solutions to $\text{RIS}(b_k, G; y_0, \boldsymbol{x}_0)$ can then be used to study the effect that various observational types have on the $k$th e.d.r. direction estimator. This will be looked at with respect to PHD in the next section.

## 4. Influence on the PHD e.d.r. space estimator

Throughout this section assume $G_\epsilon$ and $G$ are defined as in (6) with the following condition.

**Condition 4.1.** *For* $(Y, \boldsymbol{X}) \sim G$, $\boldsymbol{X} \sim N_p(\boldsymbol{\mu}, \boldsymbol{\Sigma})$.

Under Condition 4.1, let $|\lambda_1| \geq \ldots \geq |\lambda_K| > 0$ denote the absolute nonzero eigenvalues of $\overline{\text{H}}_{\boldsymbol{x}}$ that correspond to the PHD e.d.r. directions $\boldsymbol{\gamma}_1, \ldots, \boldsymbol{\gamma}_K$ and let $\boldsymbol{\Gamma} = [\boldsymbol{\gamma}_1, \ldots, \boldsymbol{\gamma}_K]$. The proof of the following Theorem can be found in the Appendix (A.1-A.3).

**Theorem 4.1.** *With notation defined above, let* $b_k^y$ *and* $b_k^r$ *denote the functionals for the kth y-based and r-based* PHD *e.d.r direction estimators such that, at G and under Condition 4.1,* $b_k^y(G) = b_k^r(G) = \boldsymbol{\gamma}_k$ *corresponds to the eigenvalue* $\lambda_k$. *Then, where* $\boldsymbol{P}_{\mathcal{S}} = \boldsymbol{\Gamma}\boldsymbol{\Gamma}^\top$,

$$\text{RIS}(b_k^y, G; y_0, \boldsymbol{x}_0) = \|(\boldsymbol{I}_p - \boldsymbol{P}_{\mathcal{S}})\boldsymbol{\Sigma}^{-\frac{1}{2}}\boldsymbol{\alpha}_{y,k}\|/|\lambda_k|,$$
$$\text{RIS}(b_k^r, G; y_0, \boldsymbol{x}_0) = \|(\boldsymbol{I}_p - \boldsymbol{P}_{\mathcal{S}})\boldsymbol{\Sigma}^{-\frac{1}{2}}\boldsymbol{\alpha}_{r,k}\|/|\lambda_k|$$

*with*

$$\boldsymbol{\alpha}_{y,k} = \left\{ (y_0 - \mu_y)\boldsymbol{\gamma}_k^\top \boldsymbol{\Sigma}^{-\frac{1}{2}} \boldsymbol{z}_0 - \lambda_k \boldsymbol{\gamma}_k^\top \boldsymbol{\Sigma}^{\frac{1}{2}} \boldsymbol{z}_0 - \boldsymbol{\gamma}_k^\top \boldsymbol{\Sigma}^{-1} \boldsymbol{\Sigma}_{\boldsymbol{x}y} \right\} \boldsymbol{z}_0 - (y_0 - \mu_y)\boldsymbol{\Sigma}^{-\frac{1}{2}} \boldsymbol{\gamma}_k,$$

$$\boldsymbol{\alpha}_{r,k} = \left\{ r_G(y_0, \boldsymbol{x}_0)\boldsymbol{\gamma}_k^\top \boldsymbol{\Sigma}^{-\frac{1}{2}} \boldsymbol{z}_0 - \lambda_k \boldsymbol{\gamma}_k^\top \boldsymbol{\Sigma}^{\frac{1}{2}} \boldsymbol{z}_0 \right\} \boldsymbol{z}_0 - r_G(y_0, \boldsymbol{x}_0)\boldsymbol{\Sigma}^{-\frac{1}{2}} \boldsymbol{\gamma}_k$$

*where* $\boldsymbol{z}_0 = \boldsymbol{\Sigma}^{-1/2}(\boldsymbol{x}_0 - \boldsymbol{\mu})$, $\boldsymbol{\Sigma}_{\boldsymbol{x}y} = \mathrm{cov}(\boldsymbol{X}, Y)$ *and* $r_G(y_0, \boldsymbol{x}_0)$ *is the* OLS *residual for* $(y_0, \boldsymbol{x}_0)$ *corresponding to the regression of* $Y$ *on* $\boldsymbol{X}$ *at* $G$.

*Remark* 4.1. The RIS measures for the PHD$_y$ and PHD$_r$ methods are equal for any given $(y_0, \boldsymbol{x}_0)$ when $\boldsymbol{\Sigma}_{\boldsymbol{x}y} = \mathbf{0}$. This can occur when $Y \perp\!\!\!\perp \boldsymbol{X}$ (a trivial case that is not supported under the assumption of $\mathrm{rank}(\overline{\mathrm{H}}_{\boldsymbol{x}}) > 0$) or for some types of link function $f$. For example, let $\boldsymbol{Z} = [Z_1, \ldots, Z_p]^\top \sim N(\mathbf{0}, \boldsymbol{I}_p)$ and suppose $Y = Z_1^2 + \varepsilon$ with $\varepsilon \perp\!\!\!\perp \boldsymbol{Z}$ and $E(\varepsilon) = 0$ then $\boldsymbol{\Sigma}_{\boldsymbol{z}y} = E(Y\boldsymbol{Z}) = \mathbf{0}$.

We now consider some examples that allow us to study the sensitivity of the PHD methods.

*Example* 4.1. Consider the multiple-index model with $E(\boldsymbol{X}) = \mathbf{0}$ and $\mathrm{cov}(\boldsymbol{X}) = \boldsymbol{I}_p$. Let $(y_0, \boldsymbol{x}_0) = (y_0, c\boldsymbol{u})$ where $c \in \mathbb{R}$ and $\boldsymbol{u} \in \mathbb{R}^p$, $\|\boldsymbol{u}\| = 1$, $\boldsymbol{u} \perp \mathcal{S}$. Then $\mathrm{RIS}(b_k^r, G; y_0, \boldsymbol{x}_0) = 0$ and $\mathrm{RIS}(b_k^y, G; y_0, \boldsymbol{x}_0) = \left| c\boldsymbol{\Sigma}_{\boldsymbol{x}y}^\top \boldsymbol{\gamma}_k / \lambda_k \right|$ for $k = 1, \ldots, K$.

This example is interesting for two reasons. Firstly, despite the fact that both PHD$_y$ and PHD$_r$ estimate the same matrix, the two methods can behave completely differently with respect to certain types of observations. Secondly, [19; 20] showed that observations of this type can be highly influential for similar dimension reduction methods such as SIR, SAVE and SIRII. However, this is not the case with PHD$_r$ so that, with respect to observations of this type, PHD$_r$ is unusual.

*Example* 4.2. Consider the single-index model

$$Y = \cos(2\boldsymbol{\beta}_1^\top \boldsymbol{X} - \pi/4) + \sigma\varepsilon$$

where $\boldsymbol{X} \sim N_p(\mathbf{0}, \boldsymbol{I}_p)$, $\varepsilon \sim N(0,1)$ and $\|\boldsymbol{\beta}_1\| = 1$. For this model $\boldsymbol{\gamma}_1 = \pm\boldsymbol{\beta}_1$ and we take, without loss of generality, $\boldsymbol{\gamma}_1 = \boldsymbol{\beta}_1$. Here, the choice of $\sigma$ is irrelevant since $\mu_y = E(Y) = E[\cos(2\boldsymbol{\beta}_1^\top \boldsymbol{X} - \pi/4)]$, $\overline{\mathrm{H}}_{\boldsymbol{x}} = E[(Y - \mu_y)\boldsymbol{X}\boldsymbol{X}^\top]$ and $\boldsymbol{\Sigma}_{\boldsymbol{x}y} = \mathrm{cov}(\boldsymbol{X}, Y) = \mathrm{cov}[\boldsymbol{X}, \cos(2\boldsymbol{\beta}_1^\top \boldsymbol{X} - \pi/4)]$ due to $E(\varepsilon) = 0$ and $\varepsilon \perp\!\!\!\perp \boldsymbol{X}$. For this model we have

$$\mu_y = \frac{1}{\sqrt{2}} e^{-2}, \quad \boldsymbol{\Sigma}_{\boldsymbol{x}y} = \sqrt{2} e^{-2} \boldsymbol{\beta}_1, \quad \lambda_1 = -\frac{2}{\sqrt{2}} e^{-2}$$

where, for verification, technical details can be found in the Appendix (A.4).

Note that, since $\|\boldsymbol{\beta}_1\| = 1$ then $\boldsymbol{\beta}_1^\top \boldsymbol{x}_0 = \|\boldsymbol{x}_0\| \cos(\theta_0)$ where $\theta_0$ is the angle between $\boldsymbol{x}_0$ and $\boldsymbol{\beta}_1$. Hence, from Theorem 4.1, we have $\mathrm{RIS}(b_1^y, G; y_0, \boldsymbol{x}_0) = c_y \|\boldsymbol{x}_0\| \sqrt{1 - \cos^2(\theta_0)}$ and $\mathrm{RIS}(b_1^r, G; y_0, \boldsymbol{x}_0) = c_r \|\boldsymbol{x}_0\| \sqrt{1 - \cos^2(\theta_0)}$ where

$$c_y = \left| \left[ (y_0 - \mu_y)\|\boldsymbol{x}_0\| \cos(\theta_0) - \lambda_1 \|\boldsymbol{x}_0\| \cos(\theta_0) - \boldsymbol{\beta}_1^\top \boldsymbol{\Sigma}_{\boldsymbol{x}y} \right] / \lambda_1 \right|,$$

$$c_r = \left| \left\{ [y_0 - \mu_y - \boldsymbol{\beta}_1^\top \boldsymbol{\Sigma}_{\boldsymbol{x}y} \|\boldsymbol{x}_0\| \cos(\theta_0)] \|\boldsymbol{x}_0\| \cos(\theta_0) - \lambda_1 \|\boldsymbol{x}_0\| \cos(\theta_0) \right\} / \lambda_1 \right|.$$

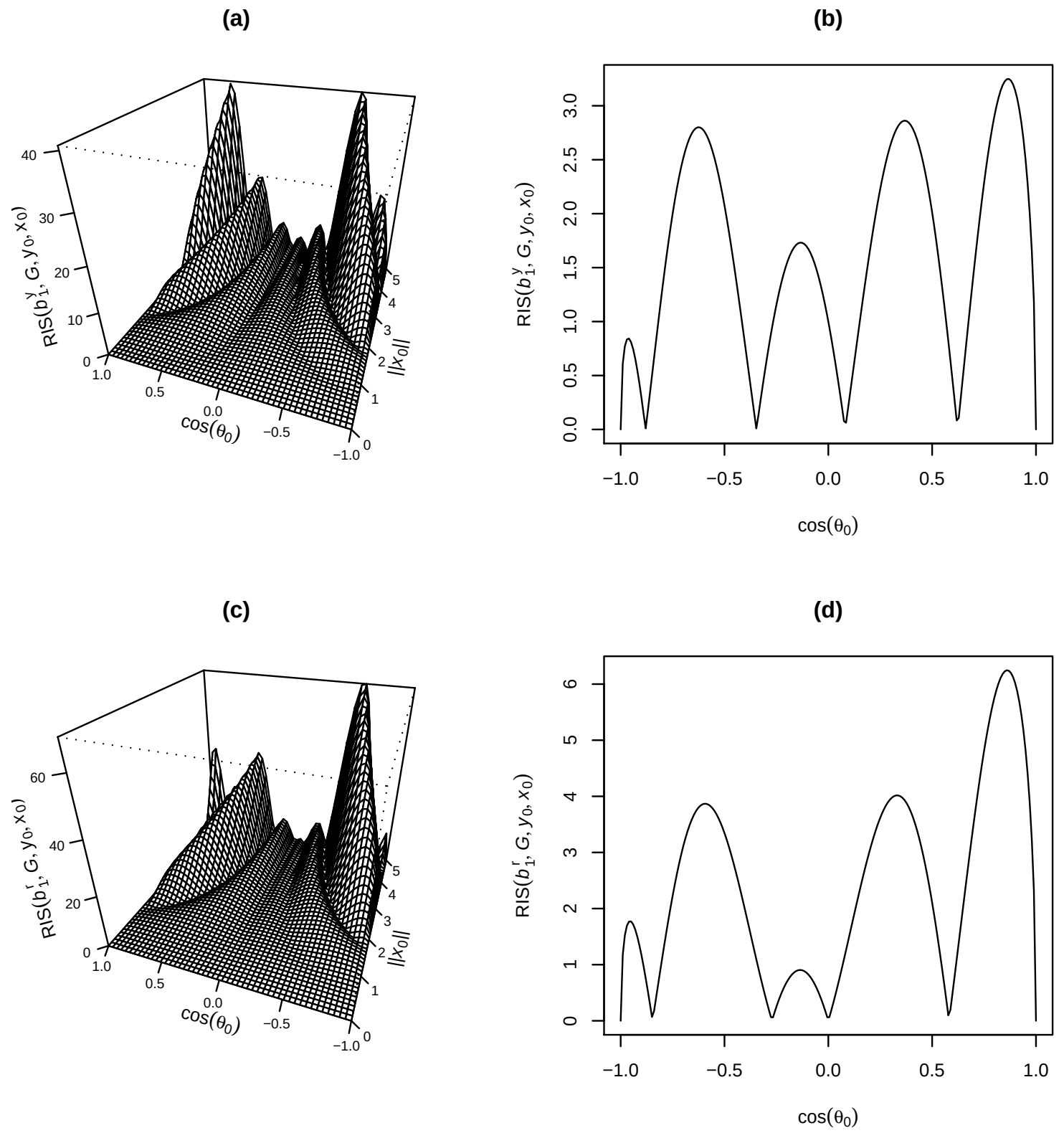

FIG 1. *Plots of (a)* $\text{RIS}(b_1^y, G; y_0, \boldsymbol{x}_0)$*, (b)* $\text{RIS}(b_1^y, G; y_0, \boldsymbol{x}_0)$ *with* $\|\boldsymbol{x}_0\| = 2$*, (c)* $\text{RIS}(b_1^r, G; y_0, \boldsymbol{x}_0)$ *and (d)* $\text{RIS}(b_1^r, G; y_0, \boldsymbol{x}_0)$ *with* $\|\boldsymbol{x}_0\| = 2$ *where* $y_0 = \cos(2\boldsymbol{\beta}_1^\top \boldsymbol{x}_0 - \pi/4)$ *for the model in Example 2;* $\|\boldsymbol{x}_0\|$ *is the length of* $\boldsymbol{x}_0$*;* $\theta_0$ *is the angle between* $\boldsymbol{x}_0$ *and* $\boldsymbol{\beta}_1$*.*

For plots of $\text{RIS}(b_1^y, G; y_0, \boldsymbol{x}_0)$ and $\text{RIS}(b_1^r, G; y_0, \boldsymbol{x}_0)$ we set $y_0 = \cos(2\boldsymbol{\beta}_1^\top \boldsymbol{x}_0 - \pi/4)$ such that $y_0|\boldsymbol{x}_0$ is consistent with the model without error. This allows us to study the sensitivity of the methods with respect to typical observations.

In Figure 1 (a) we plot $\text{RIS}(b_1^y, G; y_0, \boldsymbol{x}_0)$ for varying $\cos(\theta_0)$ and $\|\boldsymbol{x}_0\|$. It is clear from this plot that just small changes in $\theta_0$ can result in large changes of influence; in particular with increasing $\|\boldsymbol{x}_0\|$. It is also clear, however, that outliers in the predictor space, in the sense of a large $\|\boldsymbol{x}_0\|$, are not necessarily highly influential on the e.d.r. space estimator. In fact, it is possible for outlying observations to have little or no influence. In plot (b) we provide a simple cross-section of $\text{RIS}(b_1^y, G; y_0, \boldsymbol{x}_0)$ where $\|\boldsymbol{x}_0\| = 2$. This plot emphasizes the large differences in influence that can be obtained with only small rotations of $\boldsymbol{x}_0$.

Similarly, in Figure 1 (c) we plot $\text{RIS}(b_1^r, G; y_0, \boldsymbol{x}_0)$ for varying $\cos(\theta_0)$ and $\|\boldsymbol{x}_0\|$. Again it is evident that small rotations of $\boldsymbol{x}_0$ can effect large changes in influence on the r-based e.d.r. space estimator. This is again emphasized via a cross-section where $\|\boldsymbol{x}_0\| = 2$ in plot (d). For the range of $\cos(\theta_0)$ and $\|\boldsymbol{x}_0\|$ values provided here, the highest influence was achieved for the r-based method.

However, for some types of observations it is clear that this method is less sensitive than the $y$-based approach. As mentioned in Example 4.1, there is zero influence on the $r$-based e.d.r. space estimator when $\boldsymbol{x}_0 \perp \mathcal{S}$. This is again emphasized in plot (d) whereas the same observational type has non-zero influence on the $y$-based method.

## 5. Sample based sensitivity

Before we look at sample versions of the RIS we review sample versions of the influence function in general (see, for e.g., [7]). Consider a sample of $m$ observations, $w_1, \ldots, w_m$, sampled from $F$ and let $F_n$ denote the empirical distribution of this sample. Also, let $F_{n,(j)}$ denote the empirical distribution for the sample without the $j$th observation. Recall the definition of the influence function for a statistical functional $t$ given in (5). The sample influence function (SIF) for the $j$th observation on the estimator $t$ is achieved by replacing $F_\epsilon$ with $F_n$ and $F$ with $F_{n,(j)}$ such that $\text{SIF}(t, F_n; w_j) = (n-1)\{t(F_n) - t(F_{n,(j)})\}$. An approximating empirical version of the SIF can be achieved by replacing $F$ with $F_n$ in a closed-form derivation of the influence function. This approximating version is often referred to as the empirical influence function (EIF) and depends only on estimates at $F_n$ and the observation $w_j$.

### *5.1. Sample versions of the RIS*

Due to the link between the RIS and the influence function (see Remark 3.1) sample versions based on the SIF and EIF of the RIS will now be introduced to detect influential observations in practice. Let $\{(y_i, \boldsymbol{x}_i) : i = 1, \ldots, n\}$ denote a sample of $n$ observations with sample mean and covariance of the $\boldsymbol{x}_i$'s given as $\bar{\boldsymbol{x}}$ and covariance $\boldsymbol{S}$, and sample mean of the $y_i$'s given as $\bar{y}$. For this sample, let $G_n$ denote the empirical distribution and let $G_{n,(j)}$ denote the empirical distribution with the $j$th observation removed. Also, let $\widehat{\boldsymbol{\Gamma}}_y = [\hat{\boldsymbol{\gamma}}_{y,1}, \ldots, \hat{\boldsymbol{\gamma}}_{y,K}]$ denote the estimated basis for $\mathcal{S}$ at $G_n$ for $y$-based PHD and similarly denote $\widehat{\boldsymbol{\Gamma}}_r = [\hat{\boldsymbol{\gamma}}_{r,1}, \ldots, \hat{\boldsymbol{\gamma}}_{r,K}]$ for $r$-based with $\widehat{\boldsymbol{P}}_y = \widehat{\boldsymbol{\Gamma}}_y \widehat{\boldsymbol{\Gamma}}_y^\top$ and $\widehat{\boldsymbol{P}}_r = \widehat{\boldsymbol{\Gamma}}_r \widehat{\boldsymbol{\Gamma}}_r^\top$. Also suppose that $\hat{\boldsymbol{\gamma}}_{y,k}$ and $\hat{\boldsymbol{\gamma}}_{r,k}$ are associated with the eigenvalues $\hat{\lambda}_{y,k}$ and $\hat{\lambda}_{r,k}$ respectively.

Let $\theta^y_{kj}$ denote the angle between the $k$th $y$-based estimated e.d.r. direction at $G_{n,(j)}$ (i.e. without the $j$th observation) and its projection with respect to $\widehat{\boldsymbol{P}}_y$ onto the space spanned by the columns of $\widehat{\boldsymbol{\Gamma}}_y$. Then the sample RIS for the $j$th observation is

$$\text{SRIS}_{y,k}(y_j, \boldsymbol{x}_j) = (n-1)\left|\sin\left(\theta^y_{kj}\right)\right|.$$

and similarly, we define

$$\text{SRIS}_{r,k}(y_j, \boldsymbol{x}_j) = (n-1)\left|\sin\left(\theta^r_{kj}\right)\right|$$

for the $r$-based approach.

Two issues arise with the use of the SRIS. The first is that, whilst it may be employed to detect influential observations, the measure provides little interpretive information as to why an observation may or may not be influential. The second issue is that the e.d.r. space needs to be estimated $n+1$ times; once each at $G_n, G_{n,(1)}, \ldots, G_{n,(n)}$. An alternative is to approximate the SRIS by replacing $G$ with $G_n$ in the RIS to obtain a version that replaces the unknown parameters with their respective estimates at $G_n$. We will let these $y$ and $r$-based PHD empirical measures be denoted as $\text{ERIS}_{y,k}(y_j, \boldsymbol{x}_j)$ and $\text{ERIS}_{r,k}(y_j, \boldsymbol{x}_j)$ respectively.

The empirical approximations to the sample influence measures may not offer a reasonable approximation to the sample measures when $n$ is small [20]. Prendergast [20] then introduced a hybrid measure that utilized both the empirical and sample measures which improved the approximation whilst retaining the efficiency and interpretative strengths of the empirical measure. For example, from the Appendix, we have $\text{RIS}(b_k^y, G; y_0, \boldsymbol{x}_0) = \|(\boldsymbol{I}_p - \boldsymbol{P}_{\mathcal{S}})\text{IF}(\text{H}_y, G; y_0, \boldsymbol{x}_0)\gamma_k/\lambda_k\|$ where $\text{IF}(\text{H}_y, G; y_0, \boldsymbol{x}_0)$ is the influence function for the $y$-based PHD average Hessian matrix estimator. Hence the empirical RIS is, $\text{ERIS}_{y,k}(y_j, \boldsymbol{x}_j) = \|(\boldsymbol{I}_p - \widehat{\boldsymbol{P}}_y)\text{EIF}(\text{H}_y, G_n; y_j, \boldsymbol{x}_j)\hat{b}_{y,k}/\hat{\lambda}_k\|$ where $\text{EIF}(\text{H}_y, G_n; y_j, \boldsymbol{x}_j)$ is the empirical influence function for $\text{H}_y$ at $G_n$. The idea of the hybrid measure is to replace the $\text{EIF}(\text{H}_y, G_n; y_j, \boldsymbol{x}_j)$ with an efficiently computed $\text{SIF}(\text{H}_y, G_n; y_j, \boldsymbol{x}_j) = (n-1)\{\text{H}_y(G_n) - \text{H}_y(G_{n,(j)})\}$ which is derived in a closed form in terms of $(y_j, \boldsymbol{x}_j)$ and the estimates at $G_n$.

Let $\widehat{\boldsymbol{\Sigma}}_{y\boldsymbol{x}\boldsymbol{x}}$ denote the maximum likelihood estimate of $\boldsymbol{\Sigma}_{y\boldsymbol{x}\boldsymbol{x}}$ at $G_n$ and let $\boldsymbol{S}$ denote the usual unbiased estimator of $\boldsymbol{\Sigma}$ at $G_n$. Similarly, let these estimates at $G_{n,(j)}$ be denoted $\widehat{\boldsymbol{\Sigma}}_{y\boldsymbol{x}\boldsymbol{x},(j)}$ and $\boldsymbol{S}_{(j)}$ respectively. Then, for $\boldsymbol{S}_{\boldsymbol{x}y}$ denoting the usual unbiased estimate at $G_n$ for the covariance between the $\boldsymbol{x}_i$'s and $y_i$'s, it can be shown that

$$\widehat{\boldsymbol{\Sigma}}_{y\boldsymbol{x}\boldsymbol{x},(j)} = \frac{1}{n-1}\Bigg[n\widehat{\boldsymbol{\Sigma}}_{y\boldsymbol{x}\boldsymbol{x}} + \boldsymbol{S}_{\boldsymbol{x}y}(\boldsymbol{x}_j - \bar{\boldsymbol{x}})^\top + (\boldsymbol{x}_j - \bar{\boldsymbol{x}})\boldsymbol{S}_{\boldsymbol{x}y}^\top + (y_j - \bar{y})\left(\boldsymbol{I}_p - \frac{n(n+1)}{(n-1)^2}(\boldsymbol{x}_i - \bar{\boldsymbol{x}})(\boldsymbol{x}_i - \bar{\boldsymbol{x}})^\top\right)\Bigg] \tag{7}$$

which provides a closed-form solution for $\widehat{\boldsymbol{\Sigma}}_{y\boldsymbol{x}\boldsymbol{x},(j)}$. This along with the fact that (see, for example, [20])

$$\boldsymbol{S}_{(j)}^{-1} = \frac{(n-2)}{(n-1)}\boldsymbol{S}^{-1/2}\left[\boldsymbol{I}_p + \left\{\frac{(n-1)^2}{n} - \boldsymbol{z}_j^\top\boldsymbol{z}_j\right\}^{-1}\boldsymbol{z}_j\boldsymbol{z}_j^\top\right]\boldsymbol{S}^{-1/2}$$

where $\boldsymbol{z}_j = \boldsymbol{S}^{-1/2}(\boldsymbol{x}_i - \bar{\boldsymbol{x}})$, allows us to derive a closed form solution for the $\text{SIF}(\text{H}_y, G_n; y_j, \boldsymbol{x}_j)$. We will denote the hybrid measure that replaces the $\text{EIF}(\text{H}_y, G_n; y_j, \boldsymbol{x}_j)$ with this closed form solution for $\text{SIF}(\text{H}_y, G_n; y_j, \boldsymbol{x}_j)$ in the $\text{ERIS}_{y,k}(y_j, \boldsymbol{x}_j)$ as $\text{HRIS}_{y,k}(y_j, \boldsymbol{x}_j)$. Similarly we can define a version for the $r$-based approach and denote this as $\text{HRIS}_{r,k}(y_j, \boldsymbol{x}_j)$.

For comparative purposes we will also consider the Mahalanobis Distance (MD) as a measure of outlyingness for observations in the predictor space. For the $i$th observation this is given as

$$\text{MD}(\boldsymbol{x}_i) = \sqrt{(\boldsymbol{x}_i - \bar{\boldsymbol{x}})^\top S^{-1}(\boldsymbol{x}_i - \bar{\boldsymbol{x}})}.$$

We now consider an example that looks at the usefulness of these influence measures in practice.

### 5.2. Hitter's data example

The Hitter's data set, first published in Sports Illustrated (April 20, 1987), contains seventeen quantitative variables concerning regular and leading substitute hitters competing in American major league baseball in 1986. The response is the log of the salary variable where any individuals whose salary was not recorded were omitted leaving a total of $n = 263$ observations. [17] also applied PHD to this data. The three largest absolute eigenvalues for $\text{PHD}_y$ are 0.0314, 0.0238, and 0.0060 and, as such, we choose $K = 2$.

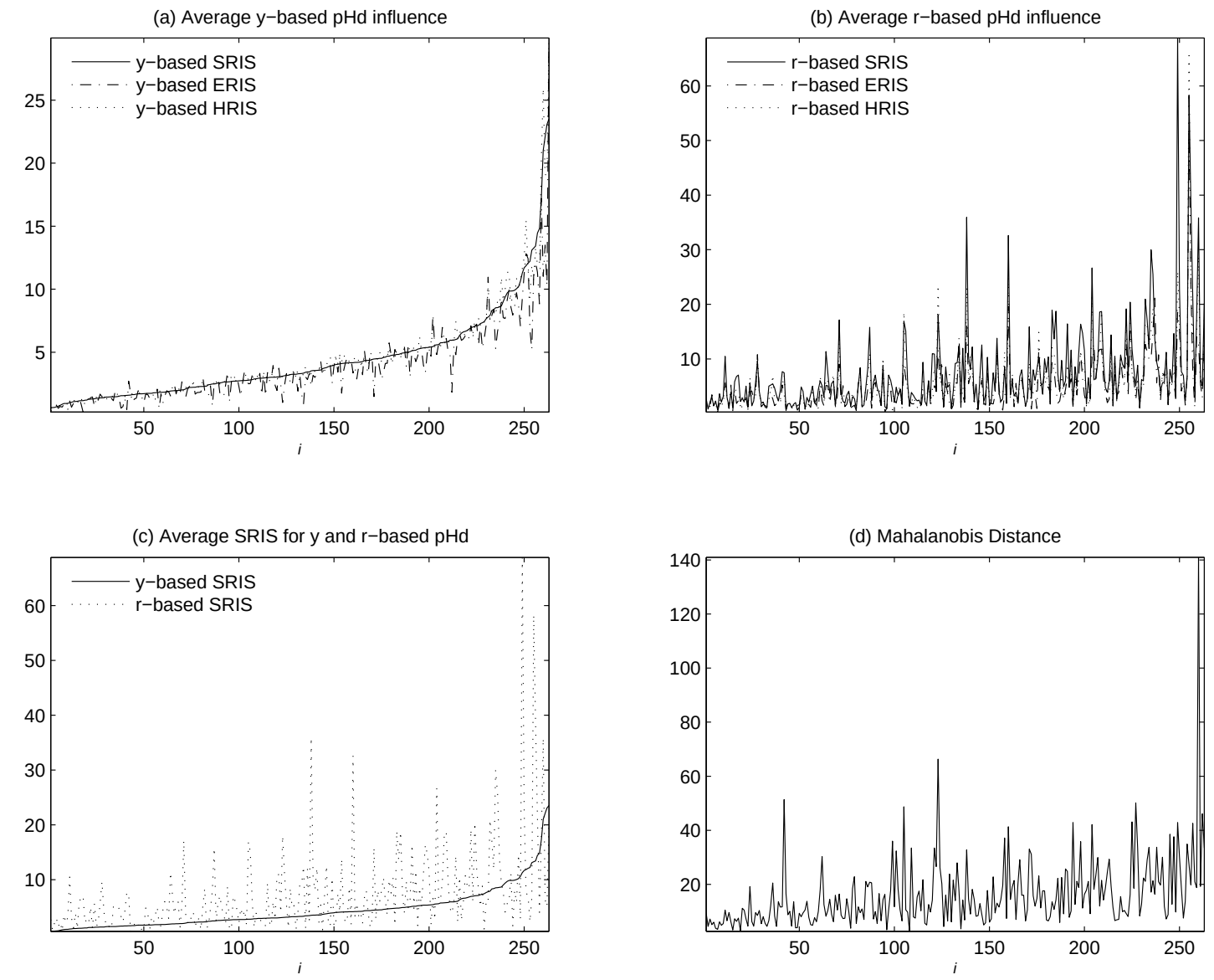

FIG 2. *Plots of (a) average* SRIS, ERIS *and* HRIS *for first two* $\text{PHD}_y$ *directions (b) average* SRIS, ERIS *and* HRIS *for first two* $\text{PHD}_r$ *directions, (c) average* SRIS *for first two* $\text{PHD}_y$ *and* $\text{PHD}_r$ *directions (d)* MD *values for the Hitter's data where i indexes the ith smallest average* SRIS *for the first two* $\text{PHD}_y$ *directions.*

In Figure 2 we provide plots of sample versions of RIS for $\text{PHD}_y$ and $\text{PHD}_r$ and the MD values for the Hitter's data. For clarity, all data in the plots are ordered according to the size of the $\text{PHD}_y$ RIS values such that $i$ indexes the $i$th smallest average of the RIS values for the 1st and 2nd $\text{PHD}_y$ directions.

Plot (a) shows that the ERIS provides a good approximation to the SRIS and can be used to successfully detect influential observations for this example with respect to $\text{PHD}_y$ however it tends to underestimate the SRIS. On the other hand the HRIS in general, gives an improved approximation for this data. Plot (b) indicates similar findings for $\text{PHD}_r$ though the ordering according to the $\text{PHD}_y$ values makes it difficult to draw direct comparisons. This will be left to the discussion of Table 1.

In Plot (c) we provide direct comparisons between the SRIS for $\text{PHD}_y$ and $\text{PHD}_r$. We see that the magnitude of influence can be significantly greater for $\text{PHD}_r$ with the largest average SRIS for $\text{PHD}_r$ being more than three-fold the largest calculated for $\text{PHD}_y$. Conversely, however, it is also clear from this plot for some observations that are highly influential on the $\text{PHD}_y$ estimator, little influence is recorded for the $\text{PHD}_r$ estimator. This plot further emphasizes the difference in the methods with regards to sensitivity.

In Plot (d) we provide the MD values for the data. Here it is evident that there is little tendency for outliers to be influential and vice versa when compared to the influence values recorded for $\text{PHD}_y$. We leave comparisons of the MD values with the influence on the $\text{PHD}_r$ estimator to the discussion of Table 1.

TABLE 1

*Spearman Rank Correlations of SRIS versus the ERIS, HRIS and MD for the Hitter's Data. Results are for the 1st estimated direction, 2nd estimated direction, and the average influence for these two directions.*

| | 1st Direction | | | 2nd Direction | | | Average Direction | | |
|---|---|---|---|---|---|---|---|---|---|
| | ERIS | HRIS | MD | ERIS | HRIS | MD | ERIS | HRIS | MD |
| $\text{PHD}_y$ | 0.898 | 0.996 | 0.435 | 0.922 | 0.992 | 0.388 | 0.935 | 0.995 | 0.506 |
| $\text{PHD}_r$ | 0.912 | 0.999 | 0.388 | 0.776 | 0.946 | 0.544 | 0.821 | 0.952 | 0.564 |

In Table 1 we provide further comparisons between the sample versions of the RIS for $\text{PHD}_y$ and $\text{PHD}_r$ using Spearman Rank Correlations.

For this example we see that the SRIS for each of the $\text{PHD}_y$ directions is approximated well by the respective ERIS values. With respect to $\text{PHD}_r$, the ERIS approximates the SRIS very well for the first direction and moderately well with respect to the second direction. The HRIS approximates the SRIS extremely well for each direction estimated using either method.

The low correlations between the SRIS and MD values emphasize that not all outliers are influential and vice versa, therefore treating them may not necessarily benefit the estimates. As such, troublesome observations from an influence perspective, may lurk within otherwise typical observations.

## 6. Conclusion

We have introduced and considered an influence measure (RIS) based on the influence function and Bénasséni's coefficient to compare two versions of Principal Hessian Directions ($\text{PHD}_y$ and $\text{PHD}_r$). Despite the fact that $\text{PHD}_y$ and $\text{PHD}_r$ seek to estimate the same Hessian matrix (and hence a basis for $\mathcal{S}$) under assumed normality of the predictor variable, we have shown that these methods can behave differently in the presence of certain observational types.

Since these differences exist in favor of either $\text{PHD}_y$ or $\text{PHD}_r$ depending on the observational types considered, we recommend the implementation of both approaches in practice and for users to give consideration to both analyses.

The unboundedness of the influence measure for both methods also reiterates the findings for other dimension reduction methods by [10; 11; 18; 19; 20] which show that such methods can fail in the presence of certain types of observations. As such, considerations for the robustification of $\text{PHD}_y$ and $\text{PHD}_r$ should be initialized.

We also provided details for how a measure such as the SRIS can be utilized at the sample level to detect influential observations in practice. Two sample measures, the ERIS and HRIS, were considered as efficient approximations to the true sample influence. The ERIS tended to underestimate the influence, in particular for small samples, though was typically successful at detecting influential observations for the example considered. For this example it is important to note that the HRIS provided an excellent approximation to the sample influence.

## Appendix A: Technical details

### *A.1. Preliminaries*

For simplicity throughout, when necessary let $\{\ldots\}^\top$ denote the transpose of the preceding term enclosed in $\{\}$. Let $T_y$ and $T$ denote the functionals for the usual mean estimators of $Y$ and $\boldsymbol{X}$ respectively where $T_y(G) = \mu_y$ and $T(G) = \boldsymbol{\mu}$. Also, let $C$ denote the function for the usual covariance matrix estimator where $C(G) = \boldsymbol{\Sigma}$ and recall that $\text{cov}_G(Y, \boldsymbol{X}) = \boldsymbol{\Sigma}_{y\boldsymbol{x}}$ with $\boldsymbol{\Sigma}_{\boldsymbol{x}y} = \boldsymbol{\Sigma}_{y\boldsymbol{x}}^\top$.

### *A.2. RIS proof for y-based PHD of Theorem 4.1*

Let $C_{y\boldsymbol{xx}}$ denote the functional defined to be, at an arbitrary distribution $(Y, \boldsymbol{X}) \sim F$ for which it exists, $C_{y\boldsymbol{xx}}(F) = \int\{Y - T_y(F)\}\{\boldsymbol{X} - T(F)\}\{\boldsymbol{X} - T(F)\}^\top dF$. At $G_\epsilon$,

$$\begin{aligned} C_{y\boldsymbol{xx}}(G_\epsilon) = &\int \{Y - T_y(G_\epsilon)\}\{\boldsymbol{X} - T(G_\epsilon)\}\{\boldsymbol{X} - T(\epsilon)\}^\top dG_\epsilon \\ =&(1-\epsilon)\boldsymbol{\Sigma}_{y\boldsymbol{xx}} + \epsilon(y_0 - \mu_y)\{(\boldsymbol{x}_0 - \boldsymbol{\mu})(\boldsymbol{x}_0 - \boldsymbol{\mu})^\top - \boldsymbol{\Sigma}\} \\ &- \epsilon(\boldsymbol{x}_0 - \boldsymbol{\mu})\boldsymbol{\Sigma}_{y\boldsymbol{x}} - \epsilon\boldsymbol{\Sigma}_{\boldsymbol{x}y}(\boldsymbol{x}_0 - \boldsymbol{\mu})^\top + O(\epsilon^2). \end{aligned} \quad (8)$$

Let $\text{H}_y$ denote the functional for the $\text{PHD}_y$ matrix estimator where $\text{H}_y(G) = \overline{\text{H}}_{\boldsymbol{x}}$ and $\text{H}_y(G_\epsilon) = \{C(G_\epsilon)\}^{-1}C_{y\boldsymbol{xx}}(G_\epsilon)\{C(G_\epsilon)\}^{-1}$. From [7], $\text{IF}(C, G; y_0, \boldsymbol{x}_0) =$ $(\boldsymbol{x}_0 - \boldsymbol{\mu})(\boldsymbol{x}_0 - \boldsymbol{\mu})^\top - \boldsymbol{\Sigma}$. Since $\{C(G_\epsilon)\}^{-1}C(G_\epsilon) = \boldsymbol{I}_p$, by way of the Product Rule we have that $[\partial \{C(G_\epsilon)\}^{-1}/\partial\epsilon]|_{\epsilon=0}\boldsymbol{\Sigma} + \boldsymbol{\Sigma}\text{IF}(C, G; y_0, \boldsymbol{x}_0) = 0$ so that

$$[\partial \{C(G_\epsilon)\}^{-1}/\partial\epsilon]|_{\epsilon=0} = -\boldsymbol{\Sigma}^{-1}\{(\boldsymbol{x}_0 - \boldsymbol{\mu})(\boldsymbol{x}_0 - \boldsymbol{\mu})^\top - \boldsymbol{\Sigma}\}\boldsymbol{\Sigma}^{-1}. \tag{9}$$

Therefore, using the Product Rule, (8) and (9),

$$\begin{aligned}\text{IF}(\text{H}_y, G; y_0, \boldsymbol{x}_0) =& \overline{\text{H}}_{\boldsymbol{x}} - \left[\boldsymbol{\Sigma}^{-1}(\boldsymbol{x}_0 - \boldsymbol{\mu})\left\{(\boldsymbol{x}_0 - \boldsymbol{\mu})^\top\overline{\text{H}}_{\boldsymbol{x}} + \boldsymbol{\Sigma}_{y\boldsymbol{x}}\boldsymbol{\Sigma}^{-1}\right\}\right] - [\ldots]^\top \\ &+ (y_0 - \mu_y)\boldsymbol{\Sigma}^{-1}\{(\boldsymbol{x}_0 - \boldsymbol{\mu})(\boldsymbol{x}_0 - \boldsymbol{\mu})^\top - \boldsymbol{\Sigma}\}\boldsymbol{\Sigma}^{-1}. \end{aligned}\tag{10}$$

Let $b_k^y$ $(k = 1, \ldots, K)$ denote the functional for the $k$th PHD$_y$ e.d.r. direction estimator where $b_k^y(G) = \boldsymbol{\gamma}_k$ and let $\theta_{k,\epsilon}^y$ denote the angle between $b_k^y(G_\epsilon)$ and $\boldsymbol{P}_{\mathcal{S}}b_k^y(G_\epsilon)$. By utilizing the identity $\sin(\theta) = \sqrt{1 - \cos^2(\theta)}$, $|\sin(\theta_{k,\epsilon}^y)| = \|(\boldsymbol{I}_p - \boldsymbol{P}_{\mathcal{S}})\{b_k^y(G_\epsilon) - \boldsymbol{\gamma}_k\}\|$ since $(I - \boldsymbol{P}_{\mathcal{S}})\boldsymbol{\gamma}_k = 0$. Therefore

$$\text{RIS}(b_k^y, G; y_0, \boldsymbol{x}_0) = \lim_{\epsilon\downarrow 0}|\sin(\theta_{k,\epsilon}^y)| = \|(\boldsymbol{I}_p - P_{\mathcal{S}})\text{IF}(b_k^y, G; y_0, \boldsymbol{x}_0)\|$$

where $\text{IF}(b_k^y, G; y_0, \boldsymbol{x}_0)$ is the influence function at $G$ for the estimator with functional $b_k^y$.

Results from [8; 9] may be used to show that the influence function for at $G$ for $b_k^y$ is (see [18])

$$\text{IF}(b_k^y, G; y_0, \boldsymbol{x}_0) = \left[\sum_{\substack{j=1\\ j\neq k}}^{K}\frac{1}{\lambda_k - \lambda_j}\boldsymbol{\gamma}_j\boldsymbol{\gamma}_j^\top + \frac{1}{\lambda_k}(\boldsymbol{I}_p - P_{\mathcal{S}})\right]\text{IF}(\text{H}_y, G; y_0, \boldsymbol{x}_0)\boldsymbol{\gamma}_k.$$

The proof is complete by noting that, from (2), $(\boldsymbol{I}_p - P_{\mathcal{S}})\boldsymbol{\Sigma}^{-1}\boldsymbol{\Sigma}_{\boldsymbol{x}y} = 0$, $(\boldsymbol{I}_p - P_{\mathcal{S}})\boldsymbol{\gamma}_k = 0$ for $k = 1, \ldots, K$ and $(\boldsymbol{I}_p - P_{\mathcal{S}})^2 = (\boldsymbol{I}_p - P_{\mathcal{S}})$.

### *A.3. RIS proof for r-based PHD of Theorem 4.1*

The same conditions and definitions as those given for the LRIS proof for PHD$_y$ are likewise employed here. Let $C_{r\boldsymbol{x}\boldsymbol{x}}$ be the functional defined at an arbitrary $F$ to be $C_{r\boldsymbol{x}\boldsymbol{x}}(F) = \int r_F(Y, \boldsymbol{X})\{\boldsymbol{X} - T(F)\}\{\boldsymbol{X} - T(F)\}^\top dF$ where $r_F(Y, \boldsymbol{X})$ denotes the OLS residual function for the regression of $Y$ on $\boldsymbol{X}$ where $(Y, \boldsymbol{X}) \sim F$ and denote $C_{r\boldsymbol{x}\boldsymbol{x}}(F) = \boldsymbol{\Sigma}_{r\boldsymbol{x}\boldsymbol{x}}$. The OLS residual functional is of the form $r_F(Y, \boldsymbol{X}) = Y - T_y(F) - \{\boldsymbol{X} - T(F)\}^\top\{C(F)\}^{-1}C_{\boldsymbol{x}y}(F)$ so that, at $G_\epsilon$,

$$\begin{aligned}C_{r\boldsymbol{x}\boldsymbol{x}}(G_\epsilon) = \quad & C_{y\boldsymbol{x}\boldsymbol{x}}(G_\epsilon) - \int\left[\{\boldsymbol{X} - T(G)\}^\top\{C(G)\}^{-1}C_{\boldsymbol{x}y}(G)\right]\{\boldsymbol{X} - T(G_\epsilon\} \\ & \times\{\boldsymbol{X} - T(G_\epsilon\}^\top dG_\epsilon.\end{aligned}$$

Then, from (8) and since $\boldsymbol{\Sigma}_{r\boldsymbol{x}\boldsymbol{x}} = \boldsymbol{\Sigma}_{y\boldsymbol{x}\boldsymbol{x}}$ when $\boldsymbol{X} \sim N_p(\boldsymbol{\mu}, \boldsymbol{\Sigma})$,

$$C_{r\boldsymbol{x}\boldsymbol{x}}(G_\epsilon) = (1 - \epsilon)\boldsymbol{\Sigma}_{r\boldsymbol{x}\boldsymbol{x}} + \epsilon\, r_G(y_0, \boldsymbol{x}_0)\{(\boldsymbol{x}_0 - \boldsymbol{\mu})(\boldsymbol{x}_0 - \boldsymbol{\mu})^\top - \boldsymbol{\Sigma}\}. \tag{11}$$

From (11), the remainder of the proof can be completed by closely following the proof for the PHD$_y$ RIS.

### *A.4. Expectation results for Example 4.2*

Firstly, recall the power series for $e^x$ given as

$$e^x = \sum_{m=0}^{\infty} \frac{x^m}{m!} = \sum_{m=1}^{\infty} \frac{x^{m-1}}{(m-1)!}. \tag{12}$$

Throughout let $Z = \boldsymbol{\beta}_1^\top \boldsymbol{X}$ where $Z \sim N(0,1)$ since $\|\boldsymbol{\beta}_1\| = 1$. The Taylor series expansion of $\cos(2Z - \pi/4)$ around $Z = \pi/8$ gives

$$\cos(2Z - \pi/4) = \sum_{n=0}^{\infty} (-1)^n \frac{2^{2n}}{(2n)!} \left( Z - \frac{\pi}{4} \right)^{2n}. \tag{13}$$

Using the moment generating function (mgf), $E[(Z - \pi/4)^{2n}] = (2n)!/(2^n n!)$ for $n \in \mathbb{N}$ so that, from (12) and (13), $E(Y) = E[\cos(2Z - \pi/4)] = \exp(-2)/\sqrt{2}$.

Since $\text{cov}(\boldsymbol{X}, Y) \in \mathcal{S}$ then $\text{cov}(\boldsymbol{X}, Y) = c\boldsymbol{\beta}_1$ for some $c \in \mathbb{R}$. Hence, $\text{cov}(Z, Y) = \boldsymbol{\beta}_1^\top \text{cov}(\boldsymbol{X}, Y)$ so that $c = \text{cov}(Z, Y)$. Using a Taylor Series expansion of $g_1(Z) = Z\cos(2Z - \pi/4)$ around $Z = 0$, we have

$$E[g_1(Z)] = \sum_{n=0}^{\infty} \frac{g_1^{(2n)}(0)}{2^n n!} \tag{14}$$

since, again via the mgf, $E[Z^{2n+1}] = 0$ and $E[Z^{2n}] = (2n)!/(2^n n!)$ for $n \in \mathbb{N}$. We also have $g_1^{(2n)}(0) = -n2^{2n}(-1)^n/\sqrt{2}$ so that, from (12) and (14), $\text{cov}(Z, Y) = \sqrt{2}\exp(-2)$.

Note that $\lambda_1 = \boldsymbol{\beta}_1^\top \overline{\text{H}}_{\boldsymbol{x}} \boldsymbol{\beta}_1 = E[(Y - \mu_y)Z^2]$ where, for $g_2(Z) = Z^2\cos(2Z - \pi/4)$, the Taylor Series Expansion around $Z = 0$ for $E(YZ^2)$ is identical to that of (14) with $g_2^{(2n)}(0)$ replacing $g_1^{(2n)}(0)$. We have $g_2^{(2n)}(0) = -n(2n - 1)2^{2n-1}(-1)^n/\sqrt{2}$ so that, from (12) and since $E(Y) = \exp(-2)/\sqrt{2}$, $E[(Y - \mu_y)Z^2] = -2\sqrt{2}\exp(-2)$.